\documentclass[twocolumn,showpacs,preprintnumbers,pre,a4paper,final,tightenlines,10pt]{revtex4}
\usepackage{graphicx}
\usepackage{amsmath}
\usepackage{multirow}
\usepackage{rotating}
\begin{document}

\title{Excitation energy transfer: Study with non-Markovian dynamics}
\author{Xian-Ting Liang\footnote{Email: xtliang@ustc.edu}}
\affiliation{Department of Physics and Institute of Modern Physics,
Ningbo University, Ningbo, 315211, China\\}

\begin{abstract}
In this paper, we investigate the non-Markovian dynamics of a model
to mimic the excitation energy transfer (EET) between chromophores
in photosynthesis systems. The numerical path integral method is
used. This method includes the non-Markovian effects of the
environmental affects and it does not need the perturbation
approximation in solving the dynamics of systems of interest. It
implies that the coherence helps the EET between chromophores
through lasting the transfer time rather than enhances the transfer
rate of the EET. In particular, the non-Markovian environment
greatly increase the efficiency of the EET in the photosynthesis
systems.

\end{abstract}

\maketitle

\section{Introduction}

Photosynthesis provides almost all the energy for life on Earth.
Therefore, many interests have been attracted on this topic in past
decades not only in biology but also in chemistry and physics
\cite{Bio-Chem-Phy}. The typical antenna of harvesting photons in
photosynthesis systems are protein complexes that hold pigments
[chlorophyll (Chl), bacteriochtorophyll (BChl), phycobilin,
carotenoid, bacteriopheophytin (BPhy), etc.]. For example, in a
green sulfur bacteria, the Fenna-Matthews-Olson (FMO) protein is a
trimer made of identical subunits, each of which contains seven BChl
molecules \cite{FMO}. Another purple bacterium, Rh. sphaeroides
\cite{Rh-sph} includes a special pair Chls (P) in the center,
accessory Bchls (B$_{L}$ and B$_{M}$) on the each side of P, and
BPhys (H$_{L}$ and H$_{M}$) next to the each BChls.

The photosynthesis starts with the absorption of photon of sunlight by the
light-harvesting pigments, followed by the excitation energy transfer (EET)
to the reaction center, where charge separation is initiated and physical
energy is converted into chemical energy \cite{PhystoChem}. At low light
intensities, the quantum efficiency of the transfer of energy is very high,
with a near unity quantum yield. Although this process has been investigated
extensively the details and the mechanism of the energy transfer has not
been fully understood today.

There is a hypothesis that the high efficiency of near unity EET
attributes to coherent superpositions between excitation states that
belong to different chromophores of the photosynthesis systems. It
is deduced from the following analogous facts. Engel et
al.\cite{Engel} found that in FMO protein the coherence between the
two excitation states clearly lasts for a longer time which is
similar to the time scale of the EET. Lee et al.\cite{Lee} also
found that in Rh. sphaeroides the coherence of the excitations
created from laser pulses on adjacent B$_{L}$ and H$_{L}$ can
prolong for longer time which is long enough for the coherent
transfer of energies between chromophores B$_{L} $ and H$_{L}$.
These kinds of experimental results imply that electronic
excitations may move coherently through these photosynthesis systems
rather than by incoherent hopping motion as has usually been
assumed.

In the aspect of theory, more than fifty years earlier,
F\"{o}rster\cite{Forster} suggested a formula of the efficiency of
the resonance EET. This theory give a picture that excitation
energies are transferred in the photosynthesis systems through
incoherent hopping from the higher levels to the lower levels.
However, this theory only describes the cases that the couplings
between chromophores are weak. The master equation of Redfield form
\cite{Redfield} has also been used in the investigations of the EET.
However, this method of the original form is limited because it
derived from the Markovian and perturbation approximations. A
modified Redfield theory, first suggested by Zhang and Fleming
\cite{Zhang-Fleming}, and later elaborated by Yang and Fleming
\cite{Yang-Fleming}, has then been used in the investigations of the
EET. It partly overcomes the shortcomings of the conventional one.
But, it cannot take into account any processes related to quantum
coherence and decoherence. Jang et al. \cite{Jang} developed the F
\"{o}rster-Dexter theory for the EET. In their method the weak
couplings between chromophores are not needed, but it asks the
couplings between the system and the environment be weak. Ishizaki
and Fleming \cite{Ishizaki-Fleming} recently used the reduced
hierarchy equation approach dealing with the quantum coherent and
incoherent dynamics of the EET. It avoids the usage of perturbative
truncation, and can describe the quantum coherent wavelike motion
and incoherent hopping in the same framework. Mohseni et al.
\cite{Mohseni} developed a quantum walk approach, based on a quantum
trajectory picture in Born-Markovian and secular approximations, as
a natural framework for incorporating quantum dynamical effects in
the EET. It is shown that the interplay between the coherent
dynamics of the system and the incoherent action of the environment
can lead to significantly greater transport efficiency than the
coherent dynamics on its own. Similar results have also been
obtained independently by Plenio et al.\cite{Plenio}. Much other
effort bas been contributed to the fields in last years \cite
{Ray-Makri,Yu-etal,Nazir,PRE65_031919_2002,MThorwart}.

However, in the existing investigations as being listed above, two
important aspects paid not enough attention to. One is how the
non-Markovian \cite{PRB78_235311_2008,JCP129_224106_2008} effects
influences the dynamics of the EET, and the other is how to describe
the evolution rate of the EET. In this paper, we shall use the
numerical path integral method developed by Makri's group
investigating the EET, where we will pay particular attention to
above two problems.

This article is organized as follows. In Sec. II, we introduce the
model and compare the numerical path integral method to the master
equation method through calculating the time evolution of the
population of sites. In Sec. III we investigate the evolutions of
the coherent term of the reduced density matrix and the transfer
rate of the EET in many kinds of conditions. It will be shown that
the coherence helps the energy transfer in the model system through
lasting the transfer time rather than increasing the transfer rate
of the EET, and as we consider the non-Markovian effects the energy
transfer will be increased greatly. Through the model study we can
obtain some important implications for energy transfer in
photosynthesis systems. Finally, Sec. IV is devoted to the
concluding remarks.

\section{Model and time evolution of the population of sites}

The Hamiltonian of the EET model can be represented by Frenkel exciton
Hamiltonian \cite{Yang-Fleming,Ishizaki-Fleming}.%
\begin{equation}
H_{tot}=H^{el}+H^{ph}+H^{re}+H^{el-ph},  \label{H-tot}
\end{equation}%
where%
\begin{eqnarray}
H^{el}&=&\sum\limits_{j=1}^{2}\left\vert j\right\rangle \varepsilon
_{j}^{0}\left\langle j\right\vert +\Delta \left( \left\vert 1\right\rangle
\left\langle 2\right\vert +\left\vert 2\right\rangle \left\langle
1\right\vert \right) ,  \label{H-el} \\
H^{ph}&=&\sum\limits_{j=1}^{2}H_{j}^{ph},  \label{H-ph} \\
H^{re}&=&\sum\limits_{j=1}^{2}\left\vert j\right\rangle \lambda
_{j}\left\langle j\right\vert ,  \label{H-re} \\
H^{el-ph}&=&\sum\limits_{j=1}^{2}H_{j}^{el-ph}=\sum%
\limits_{j=1}^{2}V_{j}u_{j}.  \label{H-el-ph}
\end{eqnarray}%
Here, $\left\vert j\right\rangle $ represents the state where only
the $j$th site is in its excitation electronic state $\left\vert
\varphi
_{je}\right\rangle $ and another is in its ground electronic state $%
\left\vert \varphi _{kg}\right\rangle $, i.e., $\left\vert j\right\rangle
\equiv \left\vert \varphi _{je}\right\rangle \left\vert \varphi
_{kg}\right\rangle $. While we define the ground state $\left\vert
0\right\rangle \equiv \left\vert \varphi _{1g}\right\rangle \left\vert
\varphi _{2g}\right\rangle $. Eq.(\ref{H-el}) is the electronic Hamiltonian
in which the Hamiltonian of the ground electronic state is set to be zero. $%
\varepsilon _{j}^{0}$ is the excitation electronic energy of the $j
$th site in the absence of phonon. $\Delta$ is the electronic
coupling between the two
sites, which responsible for the EET between the individual sites. $%
H_{j}^{ph}=\sum\nolimits_{\xi }\hbar \omega _{\xi }\left( p_{\xi }^2+q_{\xi
}^2\right) /2$ is the phonon Hamiltonian associated with the $j$th site,
where $q_{\xi },$ $p_{\xi },\ $and $\omega _{\xi }$ are the dimensionless
coordinate, conjugate momentum, and frequency of the $\xi $th phonon mode,
respectively. $\lambda _{j}=\sum\nolimits_{\xi }\hbar \omega _{\xi }d_{j\xi
}^{2}/2$ is the reorganization energy of the $j$th site, where $d_{j\xi }$
is the dimensionless displacement of the equilibrium configuration of the $%
\xi $th phonon mode between the ground and excitation states of the
$j$th site. $H^{el-ph}$ is the coupling Hamiltonian between the
$j$th site and phonon modes and the $V_{j}$ and $u_{j}$ are defined
as $V_{j}=\left\vert j\right\rangle \left\langle j\right\vert $ and
$u_{j}=\sum\nolimits_{\xi }\hbar \omega _{\xi }d_{j\xi }q_{\xi }.$
Here, we assume the phonon modes associated with one site are
uncorrelated with those of another site. If we set $\lambda
_{1}=\lambda _{2}=\lambda,$ namely, suppose that the reorganization
energy of the $1$-th and $2$-th sites are equal, and reset the zero
point energy at $(\varepsilon _{2}^{0}-\varepsilon
_{1}^{0}+u_{2}-u_{1})/2,$ and set $c_{\xi }=\left( d_{2\xi }-d_{1\xi
}\right) /2,$ we can obtain the total Hamiltonian presented with
Pauli
matrix as%
\begin{eqnarray}
H &=&\frac{\epsilon }{2}\sigma _{z}+\Delta \sigma _{x}-\sigma
_{z}\sum\nolimits_{\xi } c_{\xi }\hbar \omega _{\xi }q_{\xi }  \notag \\
&&+\sum\nolimits_{\xi }\hbar \omega _{\xi }\left( p_{\xi }^2+q_{\xi
}^2\right)/2,  \label{H-sb}
\end{eqnarray}%
where $\epsilon =\varepsilon _{2}^{0}-\varepsilon _{1}^{0}.$ Eq.(\ref{H-sb})
is the standard form of the spin-boson model. Our following investigations
start from this Hamiltonian. Thus, we can investigate the model with
numerical path integral scheme which is developed by Makri group \cite{Makri}%
, and was widely used by Thorwart group \cite{Thorwart} and many other
groups \cite{Liang} in last years.

In order to compare numerical path integral scheme to master equations
approach of Redfield form we firstly investigate the time evolution of the
population of sites for the model. The spectral distribution function is
also employed as the Drude-Lorentz density (the over-damped Brownian
oscillator model):%
\begin{equation}
J\left( \omega \right) =2\lambda \frac{\omega \gamma }{\omega ^{2}+\gamma
^{2}},  \label{J}
\end{equation}
as Ref.\cite{Ishizaki-Fleming}. The parameters are taken as the same values
as Ref.\cite{Ishizaki-Fleming}, namely, $\epsilon =100$ cm$^{-1},$ $\Delta
=100$ cm$^{-1},$ $\gamma =53.08$ cm$^{-1}$ ($\gamma ^{-1}=100$ fs)$,$ and $%
T=300$ k. To implement the numerical path integral scheme, we use
the iterative tensor multiplication (ITM) algorithm derived from the
the quasiadiabatic propagator path integral (QUAPI) technique. We
use the time step $\delta t=10$ fs which is shorter than the
correlation time of the bath and the characteristic time of the
two-level system. In order to include as much non-Markovian effect
of the bath as possible in the ITM algorithm, one should choose
$\delta k_{\max }$ as large as possible so that the dominant
non-Markovian effect of the bath is included. Here, $\delta k_{\max
}$ is the maximal number of time steps, and the $\delta k_{\max
}\delta t$ is the included memory time in the calculations. In
another word, the memory effects of the bath within time $\delta
k_{\max }\delta t$ are considered in the algorithm. If set $\delta
k_{\max }=0$, this method then reduce to the Markovian
approximation. But, the Markovian results are not same to the ones
from master equations of Redfield form, because the latter is not
only Markovian but also perturbative. The time evolutions of the
population of site 1 (suppose at the initial time the system
populated on the site 1) obtained from ITM and master equations of
Redfield form are plotted in Fig.1.

\begin{figure}[tbp]
\begin{center}
\scalebox{0.25}{\includegraphics{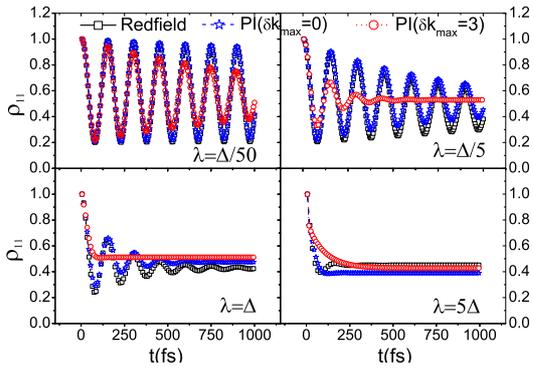}}
\end{center}
\caption{(Color online) Time evolutions of the population of site 1,
obtained from master equations of Redfield form (black line connects
black squares), and from ITM algorithm based on QUAPI of
$\protect\delta k_{max}=0$ (blue dashed line connects blue stars),
and $\protect\delta k_{max}=3$ (red dotted line connects red
circles) for different $\protect\lambda $.
Here, $\protect\epsilon =100$ cm$^{-1},$ $\Delta =100$ cm$^{-1},$ $\protect%
\gamma =53.08$ cm$^{-1}$ ($\protect\gamma ^{-1}=100$ fs)$,$ and $T=300$ k.
Initially, the system is populated on the site 1 with state $\left\vert
\protect\psi _{0}\right\rangle =\left\vert 1\right\rangle $.}
\label{fig1}
\end{figure}

In Fig.1, we plot the time evolutions of the population of site 1 by using
the master equations of Redfield form and the ITM algorithm as $%
\lambda=\Delta/50, \Delta/5, \Delta,$ and $5\Delta$. The initial state is $%
\left\vert \psi _{0}\right\rangle =\left\vert 1\right\rangle$, where $%
\left\vert 1\right\rangle$ is the basis state of the system. It is
shown that the evolutions of the population of site 1 obtained from
the master equations of Redfield form are in good agreement with the
ones obtained from the ITM algorithm of $\delta k_{\max }=0$. We
shall use $\delta k_{\max }=3,$ and $\delta t=10$ fs to investigate
the dynamics under non-Markovian approximation in this paper, so in
this case, the memory effects of the bath within $30$ fs are
included. In fact, the memory time of the bath with Drude-Lorentz
density function is longer than this value, the memory effects
beyond $30$ fs will not be included in this paper. The time
evolutions of the population of site 1 from ITM of $\delta k_{\max
}=3,$ are also plotted in Fig.1 which is quite distinct from the
Markovian results. It shows that when we investigate this kind of
system the non-Markovian effects are non-ignorable. In the
following, we shall use the ITM algorithm with $\delta k_{\max }=3$
to study the decoherence and the transfer rate of the EET. The
memory effects of the bath have not been completely included in our
calculations
but the dominating affects have been included, because when we increase $%
\delta k_{\max }$ from $3$ to $4$ the results have little change and this
change even cannot be distinguished by eyes.

\section{Time evolutions of coherence and EET}

It is reported \cite{Engel,Lee} that the decoherence time of the
coherent superposition state produced by laser pulses has the same
time scale of the EET between two adjacent chromophores. People thus
suppose that coherence may help the EET in photosynthesis. On the
other hand, the theoretical analysis shows that the environment is
not a hindrance but a help to the EET \cite{Plenio,Mohseni}. So, we
want to know further that how the coherence helps the EET and how
about non-Markovian environment compares to the Markovian one in the
helps of the EET. In this section we shall discuss these kinds of
problems through investigating above dynamical model to mimic
photosynthesis systems.

Similar to Sec.II we can easily plot the evolution of the coherence, namely,
the the evolution of the off-diagonal term in the reduced density matrix. It
is also convenient to plot the evolution of the changing rate of population
of site 2. The changing rate of the population of the site 2 in fact
reflects the transfer rate of the EET from the site 1 to the site 2.
Comparing the evolutions of the coherence and the transfer rate (denoted by $%
k$ in this paper) of the EET, we can judge whether the transfer rate
of the EET increase or not as the coherence increase, and as the
system in a non-Markovian environment instead of Markovian one.

In Figs.2 and 3 we plot the evolutions of the $\rho_{12}(t)$ and $k$
with time in different $\lambda$. Here we set $\lambda $ vary from
$\Delta/5$, to $\Delta/4, \Delta/3, \Delta/2$, and $\epsilon =100$
cm$^{-1},$ $\Delta =100$ cm$^{-1}.$ In this paper we always set
$\gamma =53.08$ cm$^{-1}$ ($\gamma ^{-1}=100$ fs)$,$ and $T=300$ k.
The initial state is set in the maximal coherent superposition state
$\left\vert \psi _{0}\right\rangle =\left( \left\vert 1\right\rangle
+\left\vert 2\right\rangle \right) /\sqrt{2}$ in Fig.2, and in the
basis state $\left\vert \psi _{0}\right\rangle =\left\vert
1\right\rangle$ in Fig.3. Figs.4 and 5 plot the evolutions of the $%
\rho_{12}(t)$ and $k$ with time t in different $\Delta$. Here, we set $%
\epsilon =100$ cm$^{-1}$, $\lambda =20$ cm$^{-1}$, and $\Delta$ vary from $%
\epsilon$, to $\epsilon/2, \epsilon/3$, and $\epsilon/4$, the initial state
is $\left\vert \psi _{0}\right\rangle =\left( \left\vert 1\right\rangle
+\left\vert 2\right\rangle \right) /\sqrt{2}$ in Fig.4, and $\left\vert \psi
_{0}\right\rangle =\left\vert 1\right\rangle$ in Fig.5. In the Figs.2-5, the
time evolutions of the coherence are plotted in the upper panels (a) and (c)
and transfer rate $k$ of the EET are plotted in the lower panels (b) and
(d). In the left panels (a) and (b) are the results of Markovian approximation $%
\delta k_{\max }=0$, and in right panels (c) and (d) are the results
of non-Markovian approximation $\delta k_{\max }=3$.

\begin{figure}[tbp]
\begin{center}
\scalebox{0.25}{\includegraphics{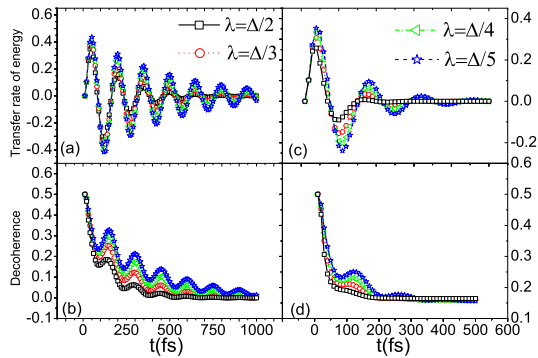}}
\end{center}
\caption{(Color online) Time evolutions of the coherence [upper
panels (a) and (c)] and transfer rate of the EET [lower panels (b)
and (d)], in Markovian approximation $\protect\delta k_{\max }=0$
[left panels (a) and (b)], and in non-Markovian approximation
$\protect\delta k_{\max }=3$ [right panels (c) and (d)], for
different values of $\protect\lambda $. The initial state is $
\left\vert \protect\psi _{0}\right\rangle =\left( \left\vert
1\right\rangle +\left\vert 2\right\rangle \right)
/\protect\sqrt{2}$. The values of other parameters are the same as
in Fig.1.} \label{fig2}
\end{figure}

\begin{figure}[h]
\begin{center}
\scalebox{0.25}{\includegraphics{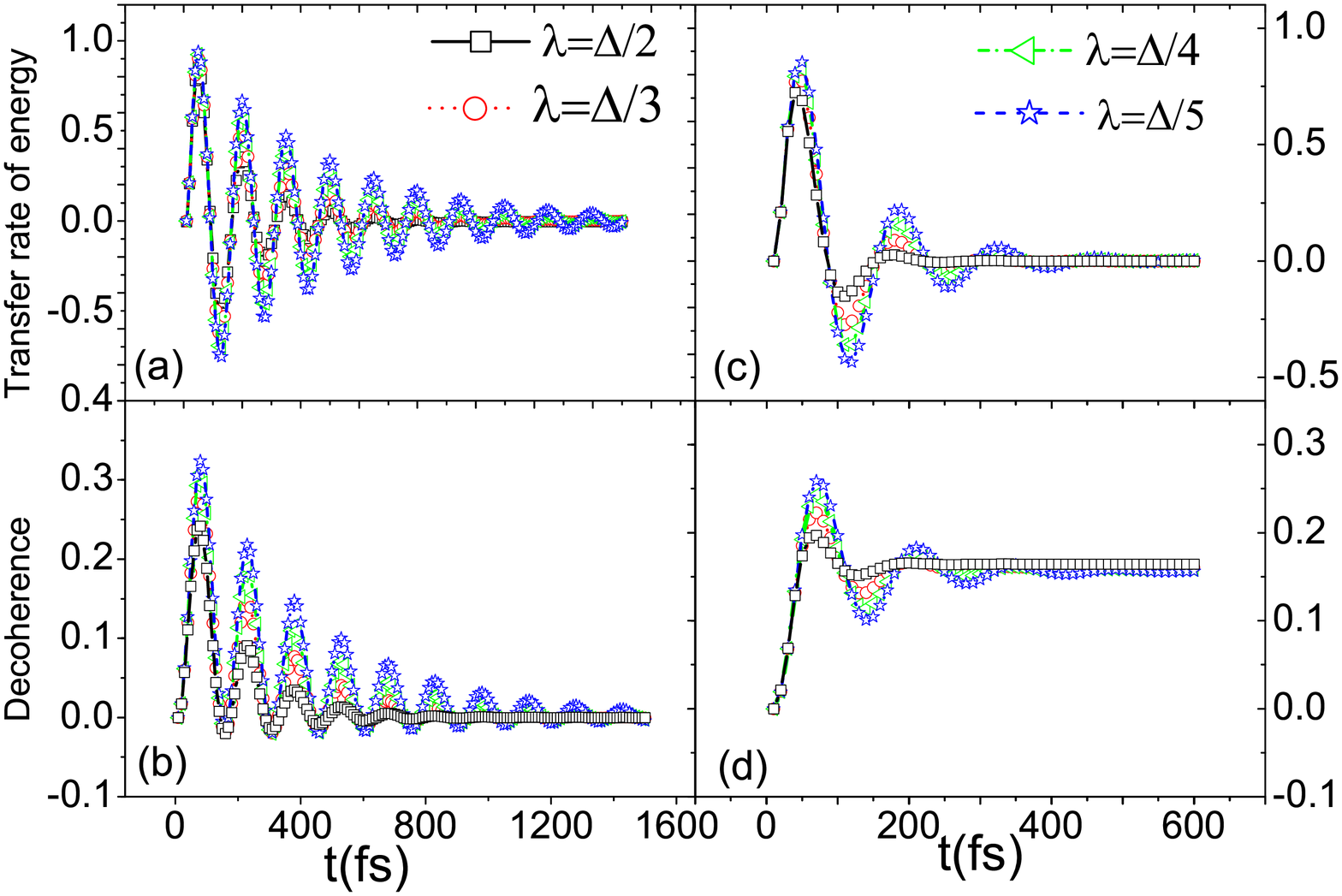}}
\end{center}
\caption{(Color online) Time evolutions of the coherence and
transfer rate of the EET in Markovian and non-Markovian
approximations. The parameters in Fig.3
are similar to them in Fig.2 except the initial state $\left\vert \protect%
\psi _{0}\right\rangle =\left\vert 1\right\rangle $.}
\label{fig3}
\end{figure}

\begin{figure}[h]
\begin{center}
\scalebox{0.25}{\includegraphics{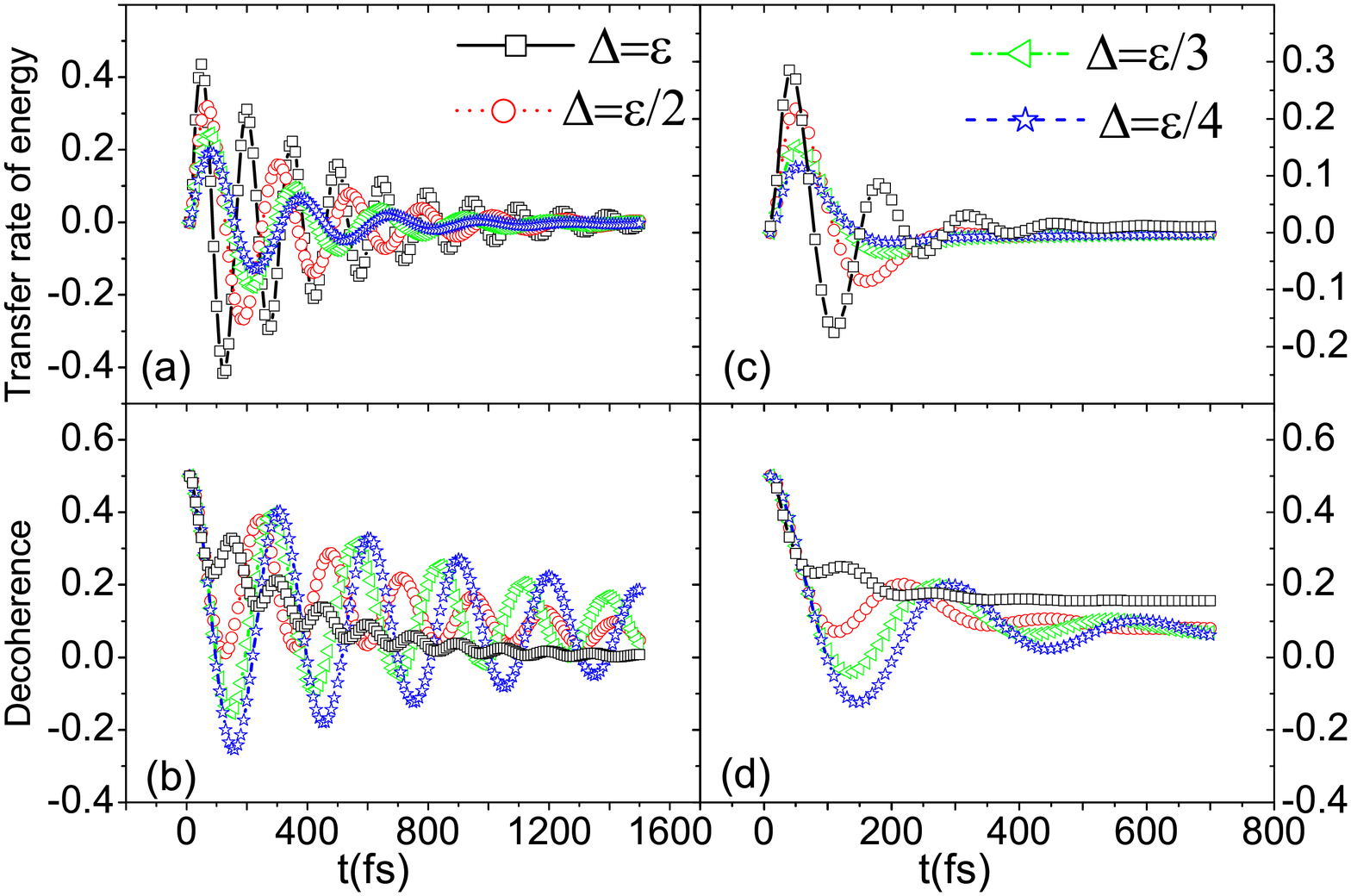}}
\end{center}
\caption{(Color online) Time evolutions of the coherence [upper
panels (a) and (c)] and transfer rate of the EET [lower panels (b)
and (d)], in Markovian approximation $\protect\delta k_{\max }=0$
[left panels (a) and (b)], and in non-Markovian approximation
$\protect\delta k_{\max }=3$ [right panels (c) and (d)], for
different values of $\Delta $. The initial state is $\left\vert
\protect\psi _{0}\right\rangle =\left( \left\vert 1\right\rangle
+\left\vert 2\right\rangle \right) /\protect\sqrt{2}$. The values of
other parameters are the same as in Fig.1.} \label{fig4}
\end{figure}

\begin{figure}[h]
\begin{center}
\scalebox{0.25}{\includegraphics{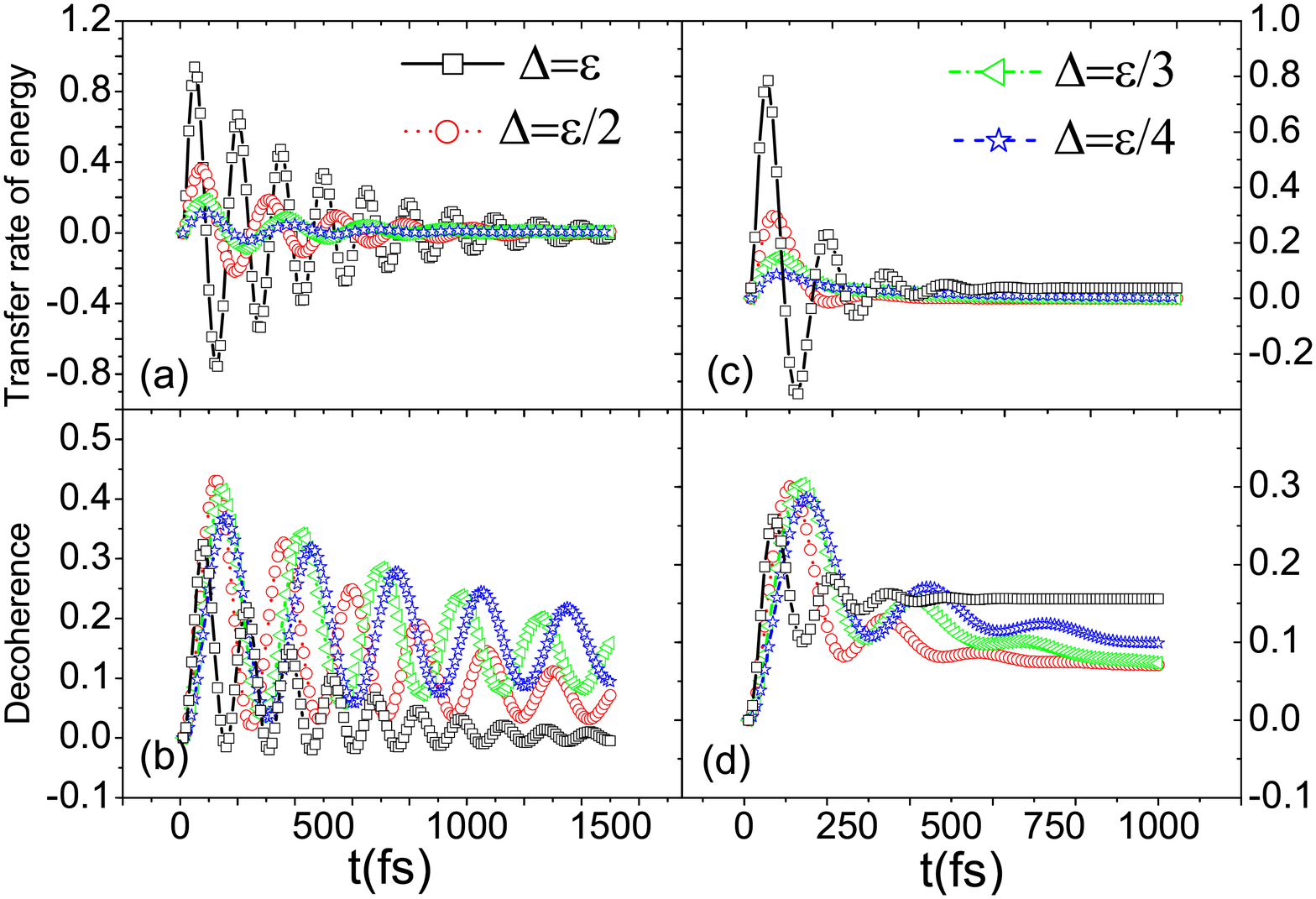}}
\end{center}
\caption{(Color online) Time evolutions of the coherence and
transfer rate of the EET in Markovian and non-Markovian
approximations. The parameters in Fig.5
are similar to them in Fig.4 except the initial state $\left\vert \protect%
\psi _{0}\right\rangle =\left\vert 1\right\rangle $.}
\label{fig5}
\end{figure}

\begin{table}[tbp]
\caption{(Obtained from Figs.2 and 3) Transfer quantity of energy and
decoherence times in different $\protect\lambda $ as $\Delta =\protect%
\epsilon =100$ cm$^{-1}$.}
\fontsize{9pt}{9pt}\selectfont
\begin{tabular}{|c|c|c|c|c|c|c|c|c|}
\hline
& \multicolumn{4}{|c|}{$\left\vert \psi _{0}\right\rangle =\left( \left\vert
1\right\rangle +\left\vert 2\right\rangle \right) /\sqrt{2}$} &
\multicolumn{4}{|c|}{$\left\vert \psi _{0}\right\rangle =\left\vert
1\right\rangle $} \\ \hline
& \multicolumn{2}{|c|}{kmax=0} & \multicolumn{2}{|c|}{kmax=3} &
\multicolumn{2}{|c|}{kmax=0} & \multicolumn{2}{|c|}{kmax=3} \\ \hline
$\lambda$ & $\tau$ & area & $\tau$ & area & $\tau$ & area & $\tau$ & area \\
\hline
$\Delta/5$ & 359 & 0.16098 & {\multirow{5}{*}{\begin{sideways}almost
same\end{sideways}}} & 0.44934 & 274 & 2.92073 & {\multirow{5}{*}{%
\begin{sideways}almost same\end{sideways}}} & 3.44756 \\
$\Delta/4$ & 265 & 0.08718 &  & 0.45430 & 390 & 2.96761 &  & 3.45388 \\
$\Delta/3$ & 207 & 0.03219 &  & 0.46212 & 260 & 2.99283 &  & 3.46209 \\
$\Delta/2$ & 149 & 0.00473 &  & 0.47830 & 240 & 2.99969 &  & 3.47830 \\
\hline
\end{tabular}%
\end{table}

\begin{table}[tbp]
\caption{(Obtained from Figs. 4 and 5) Transfer quantity of energy and
decoherence times in different $\Delta $ as $\protect\epsilon =100$ cm$^{-1}$%
, $\protect\lambda =\protect\epsilon /5$.}
\fontsize{9pt}{9pt}\selectfont
\begin{tabular}{|c|c|c|c|c|c|c|c|c|}
\hline
& \multicolumn{4}{|c|}{$\left\vert \psi _{0}\right\rangle =\left( \left\vert
1\right\rangle +\left\vert 2\right\rangle \right) /\sqrt{2}$} &
\multicolumn{4}{|c|}{$\left\vert \psi _{0}\right\rangle =\left\vert
1\right\rangle $} \\ \hline
& \multicolumn{2}{|c|}{kmax=0} & \multicolumn{2}{|c|}{kmax=3} &
\multicolumn{2}{|c|}{kmax=0} & \multicolumn{2}{|c|}{kmax=3} \\ \hline
$\Delta$ & $\tau$ & area & $\tau$ & area & $\tau$ & area & $\tau$ & area \\
\hline
$\epsilon/2$ & 410 & -0.02476 & {\multirow{5}{*}{\begin{sideways}almost same
\end{sideways}}} & 0.45031 & 800 & 2.92073 & {\multirow{5}{*}{%
\begin{sideways}almost same \end{sideways}}} & 3.45018 \\
$\epsilon/3$ & 870 & 0.09477 &  & 0.41073 & 830 & 2.90845 &  & 3.41011 \\
$\epsilon/4$ & 1320 & 0.25661 &  & 0.41204 & 1380 & 2.62069 &  & 3.37671 \\
$\epsilon/5$ & 1900 & 0.39017 &  & 0.43823 & 1700 & 2.18892 &  & 3.19348 \\
\hline
\end{tabular}%
\end{table}

The Table 1 is made by using the data of Figs.2 and 3, and the Table
2 is made from data of Figs.4 and 5. In the Tables 1 and 2 we
numerically show the decoherence times and the net area of curve $k$
in different conditions. The positive value of transfer rate of the
EET means that the energy transfers from the site 1 to the site 2,
and the negative value of the $k$ means that it transfers back to
the site 1 from the site 2. The net transfer quantity of energy from
the site 1 to the site 2 is proportional to the net \emph{area under
the curve} $k$ (where the net area denote algebraic sum of the
area). From Figs.2-5 and Tables 1 and 2 we can obtain the following
results. (1) The transfer rate of energy $k$ from the site 1 to the
site 2 are oscillations, the coherent term $\rho_{12}(t)$ are also
oscillations. The two kinds of evolution curves almost have the same
decay times. It means that the system is transferring the energy in
all of the coherence time. However, the maximum of $k$ is nether at
the maximal points nor at the minimal points of the curve
$\rho_{12}(t)$. This means that the increase of coherence does not
result in the increase of the transfer rate of the EET. Thus, we
obtain the first conclusion in this paper that coherence helps the
energy transfer through prolonging the transfer time rather than
increase the transfer rate. As seen in the figures and tables that
although the time of energy transfer is extended due to the increase
of coherence its efficiency may decrease because the energy will be
transferred between the site 1 and the site 2 back and forth. The
breath-like transfer of energy in fact is ineffective. (2) From the
Tables 1 and 2 we see that the environment really helps the energy
transfer. It is clearly shown that an environment with stronger
damping may stop the breath-like transfer of energy, which increase
the transfer quality of the EET. In particular, as the non-Markovian
effects are included the net area under the curve $k$, namely, the
transfer quality of the EET is increased greatly in all cases. It
means that the environment with non-Markovian effects will help the
EET. This is the second conclusion in this paper. It is worthwhile
to point out that as the initial state is the coherent superposition
state the energy may be transferred from site 2 to 1 then the net
area of the $k$ curve is negative as shown at 410 fs in Table 2.

\section{Summary}

To summary, in this paper we used a model introduced in
Ref.\cite{Ishizaki-Fleming} investigating the time evolutions of
population by using the numerical path integral, which is
qualitatively identical to the results obtained from reduced
hierarchy equation approach in Ref.\cite{Ishizaki-Fleming}. The path
integral method includes the non-Markovian effects of bath. By use
of the method we furthermore investigated the evolutions of the
coherence and the transfer rate of the EET between two sites. We
clearly show that there exist the energy transfer in all decoherence
time. The increase of coherence does not directly result in the
increase of the transfer rate of the EET. The helps of coherence to
the energy transfer results from the increase of transfer time
rather than the transfer rate. The energy transfer between two sites
within first several hundred femtoseconds is like a breathing motion
in Markovian environment. We have also obtained that the
non-Markovian effects play important role to the dynamics of the
energy transfer in the model systems in the initial several hundreds
femtoseconds and it results in high efficiency of the EET. Our
findings from spin-boson model study have important implications for
understanding the energy transfer in photosynthesis systems.

\begin{acknowledgements}
I like to thank Profs. Wei-Min Zhang and Yi-Zhong Zhuo for helpful
discussions. This project was sponsored by National Natural
Science Foundation of China (Grant No. 61078065), and K.C.Wong
Magna Foundation in Ningbo University.
\end{acknowledgements}

\end{document}